\documentclass[10pt,twocolumn,letterpaper]{article}

\usepackage{cvpr}              
\usepackage[table]{xcolor}
\usepackage{multirow}
\usepackage{supertabular}
\usepackage{lipsum}  
\usepackage{afterpage}
\usepackage{float}  

\usepackage{mathtools,array,dcolumn,longtable}
\definecolor{cvprblue}{rgb}{0.21,0.49,0.74}
\usepackage[pagebackref,breaklinks,colorlinks,citecolor=cvprblue]{hyperref}

\title{One-Prompt to Segment All Medical Images}

\author{Junde Wu\\
National University of Singapore\\
MBZUAI\\
University of Oxford
\and
Jiayuan Zhu \textsuperscript{\dag}\\
University of Oxford
\and
Yueming Jin \textsuperscript{\dag}\\
National University of Singapore
\and
Min Xu \\
Carnegie Mellon University\\
MBZUAI 
}

\begin{document}
\maketitle
\textbf{Apology}: We sincerely apologize to \textbf{Jiayuan Zhu}\textsuperscript{\dag} and \textbf{Yueming Jin}\textsuperscript{\dag} for their absence from the authorship list, despite their \textbf{significant contributions} to this project. Unfortunately, due to missing the CVPR registration deadline, we were unable to include them. We deeply appreciate their dedication, expertise, and insights, which have been instrumental in bringing this work to fruition. \\
\newline
\begin{abstract}
Large foundation models, known for their strong zero-shot generalization, have excelled in visual and language applications. However, applying them to medical image segmentation, a domain with diverse imaging types and target labels, remains an open challenge. Current approaches, such as adapting interactive segmentation models like Segment Anything Model (SAM), require user prompts for each sample during inference. Alternatively, transfer learning methods like few/one-shot models demand labeled samples, leading to high costs. This paper introduces a new paradigm toward the universal medical image segmentation, termed 'One-Prompt Segmentation.' One-Prompt Segmentation combines the strengths of one-shot and interactive methods. In the inference stage, with just \textbf{one prompted sample}, it can adeptly handle the unseen task in a single forward pass. We train One-Prompt Model on 64 open-source medical datasets, accompanied by the collection of over 3,000 clinician-labeled prompts. Tested on 14 previously unseen datasets, the One-Prompt Model showcases superior zero-shot segmentation capabilities, outperforming a wide range of related methods. The code and data is released as \url{https://github.com/KidsWithTokens/one-prompt}.
\end{abstract}    
\section{Introduction}
\label{sec:intro}

\begin{figure*}
    \centering
    \includegraphics[width=0.95\linewidth]{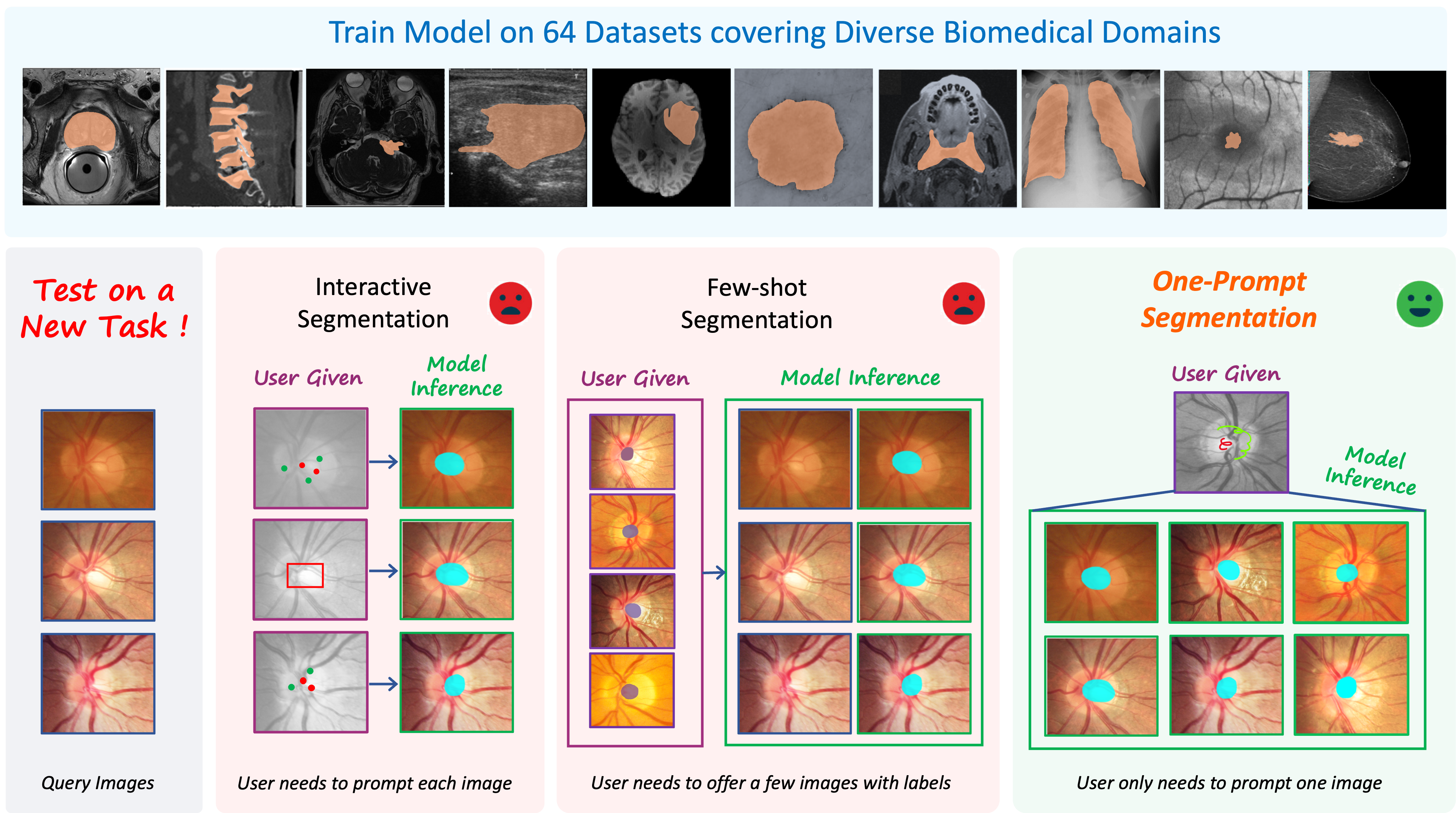}
    \caption{Medical segmentation involves a wide range of different organs, tissues and anatomies. One-Prompt Segmentation is a novel paradigm to building a foundation model that can generalize to unseen tasks. Given an unseen task, One-Prompt Model only needs the users to \textit{prompt one image} to grasp the task, which is notably cost-effective comparing with interactive and few-shot segmentation.}
    \label{fig:facial}
\end{figure*}

\begin{figure*}
    \centering
    \includegraphics[width=0.9\linewidth]{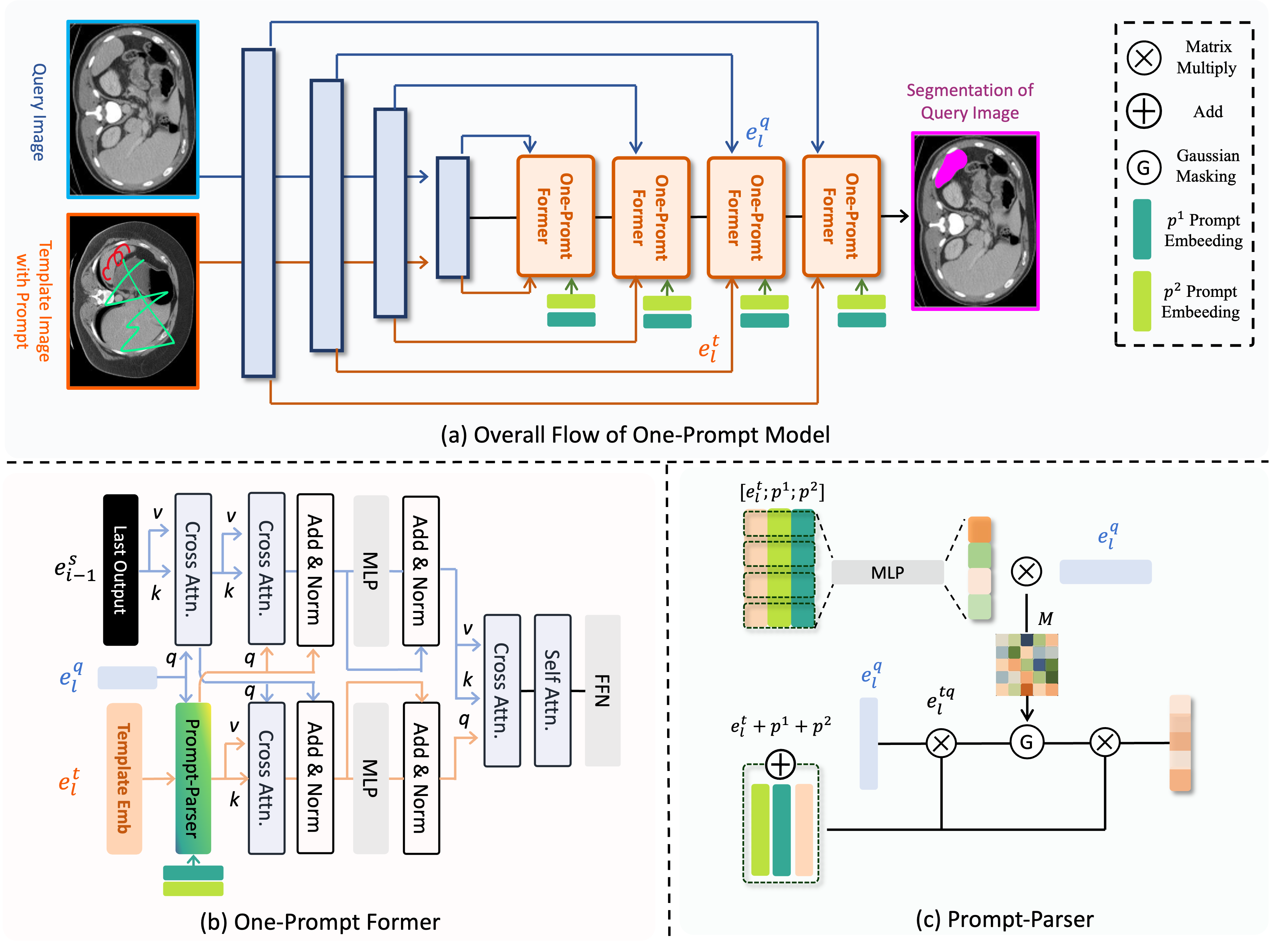}
    \caption{An illustration of One-Prompt Model, which starts from (a) an overview of the pipeline, and continues with zoomed-in diagrams
of individual Models, including (b) One-Prompt Former, and (c) Prompt-Parser.}
    \label{fig:prompt_unet}
\end{figure*}

Large foundation models pre-trained on large-scale datasets, are transforming the landscape with powerful zero-shot capabilities \cite{wei2022emergent, kirillov2023segment, wang2023seggpt, wang2023images}. These foundational models showcase an impressive ability in adapting to the tasks not seen during training. A standout example is the Segment Anything Model (SAM) \cite{kirillov2023segment}, which has gained great success for the zero-shot image segmentation. The strength of SAM lies in its interactive segmentation paradigm: the model segments the target following the user-given prompts, such as a point, a bounding box (BBox), or free text-like descriptions. 

Medical image segmentation, as a unique component of the image segmentation, plays a vital role in real-world clinical practices, including disease diagnosis and image-guided surgery. Many efforts have been made on bring this interactive foundation model to the medical image segmentation through fine-tuning \cite{wu2023medical, cheng2023sam, MedSAM}. However, most of them still need to re-training the model for each new task, leading to actually a loss of zero-shot generalization. Additionally, in these interactive models, the users have to provide prompts for each image, which is time-consuming and inapplicable for building the automatic pipeline.

Another way towards the universal medical image segmentation is few/one-shot learning \cite{butoi2023universeg, wang2019panet, peng2022hypersegnas, ouyang2020self}. In this setting, a pre-trained foundational model needs one or few \textbf{labeled} samples as the 'supportive examples', to grasp a new specific task. However, securing labels for new tasks is not always feasible. Furthermore, the success of these methods heavily depends on the number of supportive examples provided. For example, UniverSeg\cite{butoi2023universeg} achieves competitive performance with 64 supportive samples. But obtaining such amount of data can be challenging in real clinical practice.

In this paper, we introduce a new paradigm for the universal medical image segmentation, called \textit{One-Prompt Medical Image Segmentation}. This method combines the strengths of both one-shot and interactive models to meet the real clinical requirements. Specifically, given an unseen task, the user only needs to provide \textit{one prompted sample} to the trained model, then it can perform well at this new task without any retraining or fine-tuning, even for tasks significantly different from those encountered during training. An illustration is shown in Fig. \ref{fig:facial}. 

The success of the One-Prompt Model is driven by three key factors. First, we propose novel One-Prompt Former modules as the model decoder. Such design helps to efficiently integrate the prompted template feature into the process of query image segmentation. Secondly, we gather a large-scale data collection comprising 78 open-source datasets covering diverse biomedical domains. Our model is trained on 64 datasets, with clinicians prompting a part of the data. Moreover, to further enhance the clinical utility, we offer four different prompt types, which are \textit{Click}, \textit{BBox}, \textit{Doodle}, and \textit{SegLab}. The \textit{Click} and \textit{BBox} work the same as those in SAM \cite{kirillov2023segment}. \textit{Doodle} allows the users to freely draw on the image, helpful for prompting irregular organs like the pancreas or lymph glands. \textit{SegLab} allows the user to provide the segmentation mask as the prompt for representing more detailed tissues like vessels.

In sum, our contributions can be summarized as follows:
\begin{itemize}
\item We introduce novel One-Prompt Segmentation, which is a strong but low-cost paradigm for the universal medical image segmentation. 

\item We propose a model with unique One-Prompt Former to fuse the prompted template feature with the query feature in the multiple feature scales.

\item We set four different prompt types for better prompting the special medical targets, thus to meet the various clinical practices .

\item We gather a large-scale collection comprising 78 open-source datasets to train and test the model, and also annotated over 3000 of the samples with clinician-given prompts.

\end{itemize}
\section{Method}\label{sec:formatting}
Consider a set $D$ containing all medical image segmentation tasks. Each task $d$ consists of image-label pairs ${x^{d}, y^{d}}$. In conventional fully-supervised segmentation methods, a function $y^{d} = f^{d}_{\theta}(x^{d})$ is typically learned to estimate a segmentation map $y^{d}$ based on an input image $x^{d}$. However, this function $f^{d}_{\theta}$ is tailored exclusively for the specific task $d$. In the case of few-shot strategies, the target is to learn a universal function $y^{d} = f_{\theta}(x^{d},S^{d})$ performing on any task $d$ guided by the task-specific support set $S^{d} = \{(x^{d}_{j}, y_{j}^{d})\}^{n}_{j=1}$, comprising example image \& label pairs available for task $d$.

In contrast, our One-Prompt Segmentation learns a more general function $y = f_{\theta}(x^{d}_{j}, k^{d})$ performing on any task $d$, where $k^{d} = \{ x^{d}_{c}, p^{d}_{c} \}$ comprising one fixed template image $x^{d}_{c}$ and a paired prompt $p^{d}_{c}$ available for task $d$. This prompt can be freely chosen by the users from different types like \textit{Click}, \textit{BBox}, \textit{Doodle}, and \textit{SegLabel}. This learning paradigm is more user-friendly in the clinical practice in that users only need to provide a single sample with prompts, and the model can adapt to any new task in one forward pass. This makes it easy for clinicians without a computer science background to use the system without the complexities of training or fine-tuning.

It is worth noting that interactive and one-shot models can be seen as special cases of One-prompt Segmentation. In specific, when $x^{d}_{c} = x^{d}$, it works as an interactive segmentation model, and when $p^{d}_{c}$ is a segmentation label, it aligns with a one-shot model.

\subsection{Prompts}\label{AA}

Our model supports four types of prompts. Going beyond the usual \textit{Click} and \textit{BBox}, users can also use segmentation labels (\textit{SegLab}) and freehand doodles (\textit{Doodle}) as prompts. Each prompt type is best suited for its specific situations. For example, \textit{Click} work well for obvious lesions like melanoma. \textit{BBox} are effective for lesions with blurry boundaries but can be refined well by the boxes, such as the optic cup. \textit{SegLab} are ideal for scenarios with detailed features, like complex vessels. \textit{Doodle} are handy for organs with varies and unusual structures, like the pancreas and mandible. An illustration of it is shown in Section C. in the supplementary.

All prompts are represented using two embeddings, denoted as $p^{1}$ and $p^{2}$. For \textit{Click} and \textit{Doodle}, the two embeddings are used to denote the foreground and background. \textit{BBox} use them to denote the left-top and right-bottom corner points. For these three kinds of prompts, we use positional encoding to compress the coordinate information of the prompts, then add them to the learnable embeddings. These embeddings learn themselves the concepts they to represent. \textit{SegLab} is converted into an embedding using a pre-trained autoencoder. Two prompt embeddings share the same parameters in this case.

\subsection{Model}

The One-Prompt Model comprises an image encoder and a sequence of One-Prompt Former as the decoder, illustrated in Fig. \ref{fig:prompt_unet} (a). The model takes three inputs: the query image $x_{q}$, the template image $x_{t}$, and the prompt of the template image $p_{t}$, and subsequently predicts the segmentation of the query denoted as $y_{q}$. The multi-scale features of the encoder and decoder are skip-connected.

\noindent \textbf{Encoder}
The image encoder can be CNN based \cite{isensee2021nnu} or ViT based \cite{chen2021transunet}. We show a CNN based encoder in the figure for the simple illustration. The query sample $x_{q}$ and the template sample $x_{t}$ will both go through the same encoder to get the feature $f_{q}$ and $f_{t}$. 

\noindent \textbf{One-prompt Former}
We then decode the down-sampled query feature by incorporating the prompt embeddings, multi-scale template features, and multi-scale query features together through a sequence of our proposed One-Prompt Former. All feature maps are pachlized, flattened and projected to the embedding $e \in \mathcal{R}^{N \times L}$ for further processing. The One-Prompt Former primarily consists of attention blocks, and its structure involves two parallel branches for processing query and template features, as depicted in Fig. \ref{fig:prompt_unet} (b). 

In each One-Prompt Former block (considering the $i^{th}$ block), the Cross Attention \cite{chen2021crossvit} in the query branch initially takes the $l^{th}$ level skip-connected query embedding $e_{l}^{q}$ as the \textit{query} and the last output embedding $e_{i-1}^{s}$ serves as both the \textit{key}  and \textit{value}, followed by another Cross Attention to incorporate the template feature symmetrically. Simultaneously, the template branch employs a proposed Prompt-Parser to integrate the prompts $p$ with $e_{l}^{q}$ and $e_{l}^{t}$, followed by another Cross Attention to integrate the query feature. In the end, a Cross Attention integrates the two branches by transferring the prompted template segmentation to the query domain. Then a self-attention followed by Feedforward Neural Network (FNN) are employed to project the embedding to the desired length.

\noindent \textbf{Prompt-Parser}\label{AA}
In the template branch, we propose a simple Prompt-Parser to mix the prompt, query and template feature in an effective way. We show an illustration of the Prompt-Parser in Fig. \ref{fig:prompt_unet} (c). The high-level idea is to produce an adaptive attentive mask $M$ to activate the prompted target on the query-template-integrated embedding $e_{l}^{tq}$:
\begin{equation}
e_{l}^{tq} = e_{l}^{t} (p^{1} + p^{2} + e_{l}^{q}).
\end{equation}
We then divide Prompt-Parser to a Prompting Step and a Masking Step. In the Prompting Step, we build a mask $M$ adaptive to the different feature scale by mixing prompts with the given $e_{l}^{t}$ and $ e_{l}^{q}$. Specifically, we fist apply a MLP layer on the stacked embedding $[f^{t};p^{1};p^{2}] \in \mathcal{R}^{3N \times L}$. 
An MLP layer with weight $w \in \mathcal{R}^{N \times 3N} $ is applied along $N$ to mix three different embeddings and reduce its dimension back to $N$. $e_{l}^{q}$ is then matrix multiply on it to transfer its activation to the query domain. The process can be formalized as:
\begin{equation}
M =  w [e_{l}^{t};p^{1};p^{2}] (e_{l}^{q})^{T}. 
\end{equation}
Then in the Masking Step, we apply $M \in \mathcal{R}^{N \times N}$ to $e_{l}^{tq}$ through a proposed \textit{Gaussian Masking} operation:
\begin{equation}
e^{G} =  Max (e_{l}^{tq} * k_{G}[Conv(M)], e_{l}^{tq}),
\end{equation}
where $k_{G}$ is Gaussian kernel, * denotes general convolution operation. \textit{Gaussian Masking} first projects $M$ to a 2-channel feature map by convolution layer. Then we generate $k_{G}$ by taking two channels as mean and variance respectively. $k_{G}$ then multiply with $e_{l}^{tq}$ in a pixel-wise manner to enlarge the prompted space but with uncertainty. Finally, we select the maximum value between the original feature and the smoothed one, preserving the highest activation and eliminating low-confidence uncertainty regions. The output is obtained by finally multiplying with $e_{l}^{t}$.

\subsection{Training and Loss}
We divided our One-Prompt dataset into 64 datasets for model training and 14 datasets for testing. Each training dataset is further split into three parts: a prompted template split, a training split, and a validation split. Similarly, each test dataset has a test split and a prompted template split. Human users prompt each sample in the template split, for both training and testing. Model training is performed on the template and training splits across all datasets. In each iteration, we randomly pick one prompted template from the template set of the same dataset with the query image. Training is collectively conducted across the 64 datasets. Our final loss is a simple sum of cross-entropy loss and dice loss.

During the inference stage, we randomly choose a prompted template from the template split and run the model over the test/validation split for the evaluation. Unless otherwise specified, we run the model 50 times and ensemble the predictions to mitigate variance.
\section{One-Prompt Data}
\subsection{Data Source}
In order to construct a foundation model with high generalization on the unseen tasks, we train our model on large-scale and diverse medical images consisted by online open-access datasets. Our data source is constructed from 78 datasets encompassing diverse medical domains and imaging modalities. The dataset covers a wide array of organs, such as lungs \cite{saporta2021deep, setio2017validation, simpson2019large}, eyes \cite{fang2022refuge2,stare, ma2020rose, orlando2020refuge}, brain \cite{baid2021rsna, gollub2013mcic, hernandez2022isles, kuijf2019standardized, kuklisova2011dynamic}, and abdominal \cite{ji2022amos, bloch2015nci, heller2021state, kavur2021chaos, lambert2020segthor, landman2015miccai, lemaitre2015computer, litjens2014evaluation, luo2021word, ma2021abdomenct, radau2009evaluation, simpson2019large}. A detailed list of One-Prompt datasets is released with our code.

\subsection{Prompt Annotation}
A team of clinicians prompt over 3000 samples across over all collected dataset. These samples are meticulously selected by experienced annotators to ensure diversity and comprehensiveness. The clinicians involved in prompting come from diverse backgrounds, including cardiologists, dermatologists, gastroenterologists, neurologists, oncologists, pulmonologists, rheumatologists, endocrinologists, and ophthalmologists. They are encouraged to choose datasets aligned with their expertise during the prompting process.

In this process, all four prompt types are available for each sample, and clinicians are encouraged to use the most convenient prompt tool for the given targets. The clinicians employ a browser-based interactive segmentation tool to prompt images. Upon prompting, ground-truth masks immediately appear on the images based on the given prompts. Clinicians have the flexibility to refine their prompts, but adjustments are suggested only if they feel their initial prompt was incorrect.
Our prompt-based segmentation operates in real-time directly within a browser. Notably, we do not set strict constraints on prompt quality. Clinicians are encouraged to prompt images in their most natural way, with a suggested time limit of no more than 5 seconds for each image. The prompting details on the test set can be found in Section \ref{sec:Human-User Prompted Evaluation}.

\section{Experiments}
In this section, we present task generalization experiments with One-Prompt Model. We divide out 14 tasks in our available datasets, which differ significantly in term of image modalities and target structures, as our held-out test set. This set comprises 8 MICCAI2023 Challenge tasks, encompassing various anatomies including kidney tumor \cite{heller2023kits21}, liver tumor \cite{quinton2023tumour}, breast cancer \cite{tdsc-abus2023}, nasopharynx cancer \cite{astaraki2023fully}, vestibular schwannoma \cite{CrossMoDA23}, mediastinal lymph node \cite{LNQ2023}, cerebral artery \cite{CAS2023}, and inferior alveolar nerve \cite{ToothFairy}. The other 6 tasks including the segmentation of white blood cell \cite{WBC}, optic cup \cite{fang2022refuge2}, mandible \cite{pendal}, coronary artery \cite{CadVidSet}, pancreas \cite{Pancreasdata}, and retinal blood vessel \cite{stare}. We assess the model performance on each test dataset using a specific prompt type, informed by the observation that users tend to favor specific prompts for particular tasks. We provide our implementation and data processing details in the appendix.

\subsection{Human-User Prompted Evaluation}\label{sec:Human-User Prompted Evaluation}
For the evaluation, we involved human users to simulate real-world interactions for prompt-based segmentation. We assigned 15 users to prompt about 10\% of the test images. The users comprised 5 regular individuals with a clear understanding of the task but no clinical background, 7 junior clinicians, and 3 senior clinicians. This aims to simulate real-world prompting scenarios such as clinical education or semi-automatic annotation.

\begin{figure*}
    \centering
    \includegraphics[width=\linewidth]{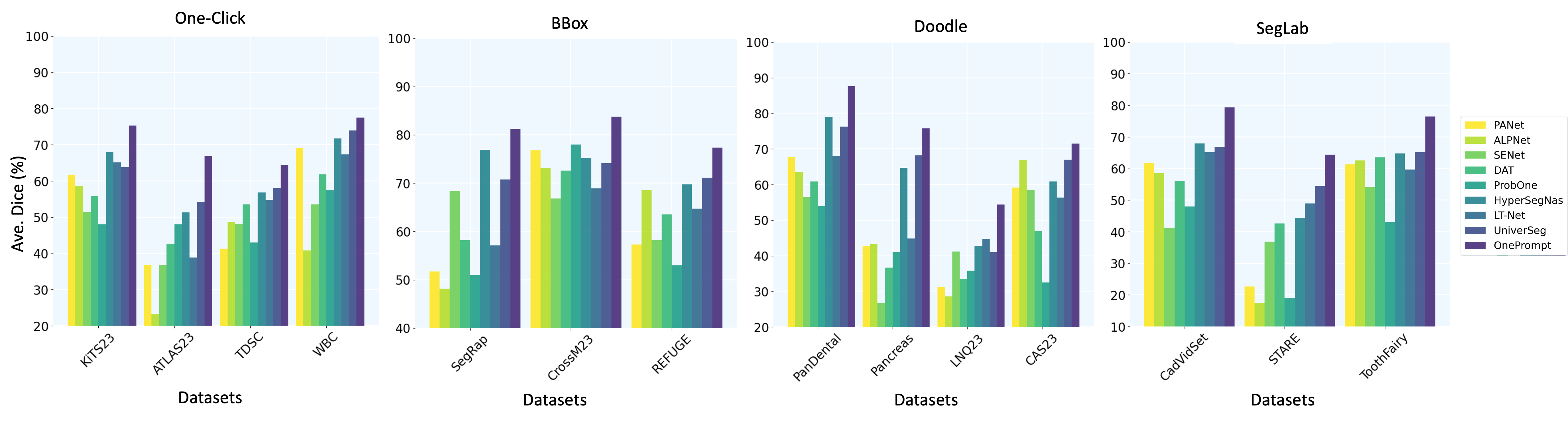}
    \caption{One-Prompt Model v.s. Few/One-shot Models on 14 held-out test datasets with 4 different prompts.}
    \label{fig:trans_pkg}
\end{figure*}

\subsection{One-Prompt Transfer Capability} \label{sec:exp-oneprompt}
\noindent \textbf{Task} 
We first validate the One-Prompt transfer capability by comparing it with few/one-shot learning baseline using varies prompts. Our main objective is to assess the generalization of One-Prompt Model in solving unseen tasks. 
We compare with various few-shot models: PANet\cite{wang2019panet}, ALPNet\cite{ouyang2020self}, SENet\cite{roy2020squeeze}, UniverSeg\cite{butoi2023universeg}, all provide with the same template to run. Few-shot methods are all given only one template in the testing for the fair comparison. Additionally, we compare with one-shot models: DAT\cite{zhao2019data}, ProbOne \cite{ding2021modeling}, HyperSegNas\cite{peng2022hypersegnas}, and LT-Net \cite{wang2020lt}.  
All these models are trained on the same dataset as ours, and are all given segmentation labels as the 'prompt' as they could not accept sparse prompts.

\noindent \textbf{Data}
In this comparison, we conduct tests on the held-out test set with 14 different tasks. Among them, KiTS23\cite{heller2023kits21}, ATLAS23\cite{quinton2023tumour}, TDSC\cite{tdsc-abus2023}, and WBC\cite{WBC} datasets using the one \textit{Click} prompt. For the SegRap \cite{astaraki2023fully}, CrossM23 \cite{CrossMoDA23}, and REFUGE \cite{fang2022refuge2} datasets, we employ the \textit{BBox} Prompt. The \textit{Doodle} prompt is applied to the Pendal \cite{pendal}, Pancreas-CT \cite{Pancreasdata}, LNQ23 \cite{LNQ2023}, and CAS23 \cite{CAS2023} datasets, while the \textit{SegLab} prompt is used for CadVidSet \cite{CadVidSet}, STAR \cite{stare}, and ToothFairy \cite{ToothFairy} datasets

\begin{figure*}
    \centering
    \includegraphics[width=\linewidth]{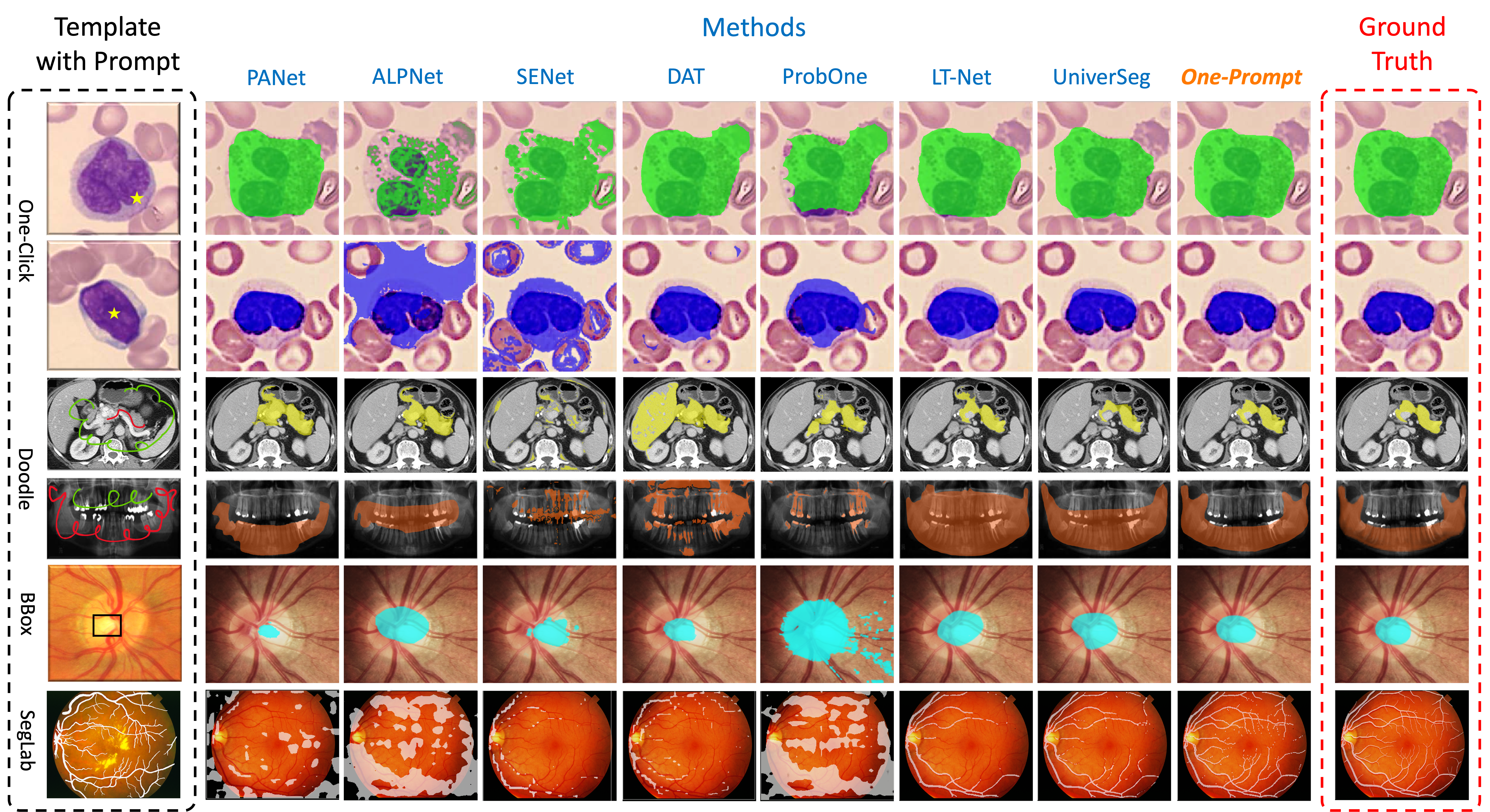}
    \caption{Visualized comparison of One-Prompt Model and few/zero-shot models. One-Prompt Model is given templates with prompts for the prediction. Few/zero-shot models are given templates with segmentation labels for the prediction.}
    \label{fig:show_res}
\end{figure*}

\noindent \textbf{Results}
Fig. \ref{fig:trans_pkg} illustrates the average Dice score per task for each method, and Fig. \ref{fig:show_res} provides the comparison on visualized results. It is worth noting that the compared few/one-shot models all necessitate the segmentation label as the 'prompt.' Therefore, they have a comparative advantage in contrast to our model. Despite this, our model consistently outperforms the competitors by significant margins, showcasing its robust generation ability across various tasks. In a fair comparison where all methods are provided segmentation labels (Fig. \ref{fig:trans_pkg} SegLab), our model demonstrates more substantial leads, which averagely outperforms the second 11.2\%, which is the most among all the prompt settings.

\begin{figure}
    \centering
    \includegraphics[width=\linewidth]{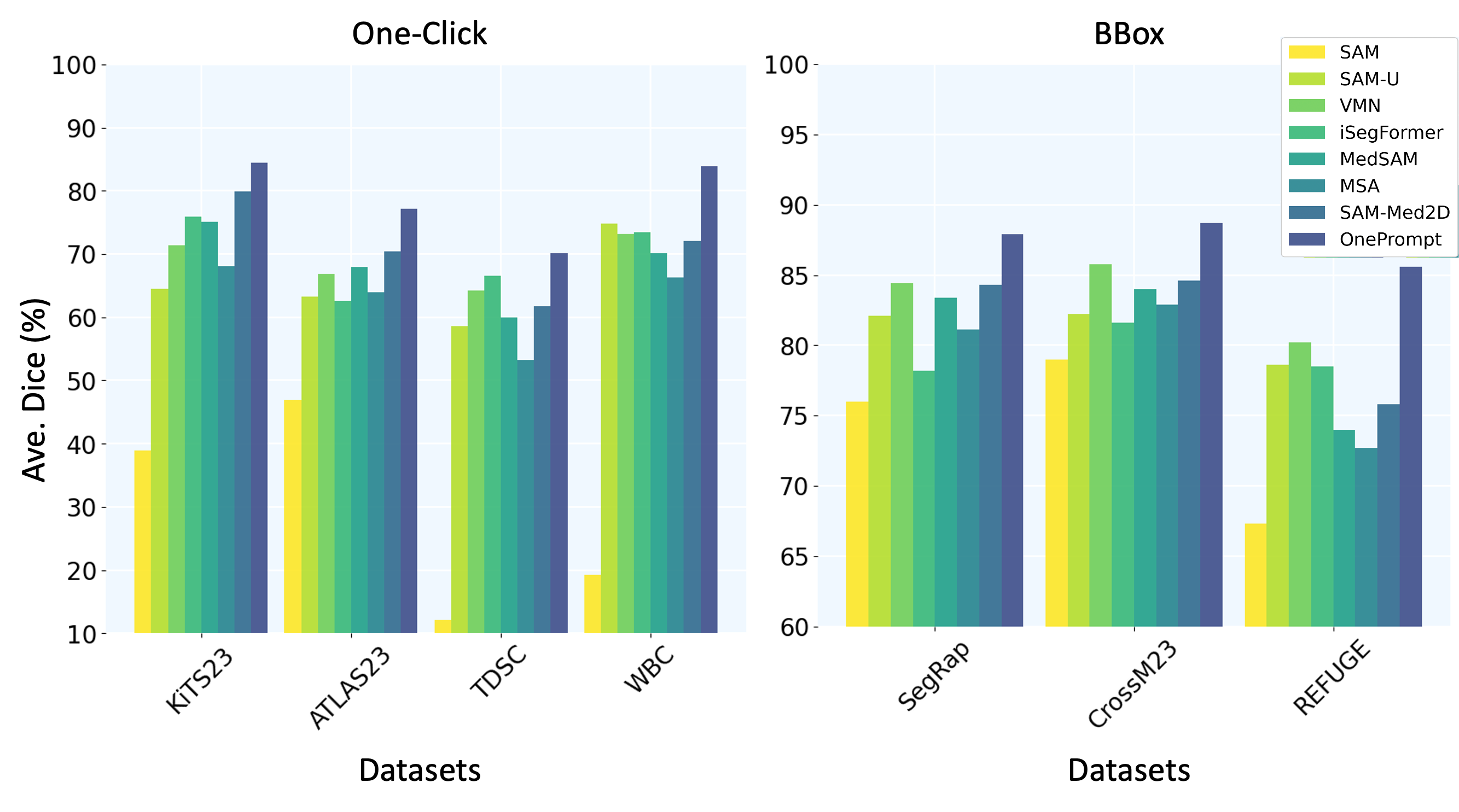}
    \caption{One-Prompt Model v.s. Interactive Segmentation Models on 7 held-out datasets with \textit{One-Click} and \textit{BBox} prompts. }
    \label{fig:inter_pkg}
\end{figure}

\subsection{Interactive Segmentation Capability}
\noindent \textbf{Task} \label{sec:exp-interactive}

Interactive segmentation models achieve zero-shot generalization by prompting each of the test sample. When we offer the One-Prompt Model with the same query image and prompted template image, the model degrades to a standard interactive segmentation model. We compare this setting with other interactive segmentation models, including vanilla SAM \cite{kirillov2023segment}, SAM-U \cite{deng2023sam}, VMN \cite{zhou2023volumetric}, iSegFormer\cite{liu2022isegformer}, MedSAM \cite{ma2023segment}, MSA \cite{wu2023medical}, and SAM-Med2D \cite{cheng2023sam}. Except vanilla SAM, all models are trained on the same dataset as ours. Since most of these models only accept \textit{Click} and \textit{BBox} prompts, \textit{Doodle} and \textit{SegLab} prompt settings are not included in this comparison. Since all these models need the prompt on each input image, we simulate the \textit{oracle} prompts (details in Section \ref{sec:exp-ablation}: Effect of prompt quality \& types in the inference) over the images if needed. It is worth noting that we did not re-train One-Prompt Model on the simulated prompts: we use the same trained One-Prompt Model as that in the last section, but only offered the simulated prompts in testing for the possible of comparison.

\noindent \textbf{Data} We conduct the comparison on all 14 held-out test datasets.

\noindent \textbf{Results}.
We present a quantitative comparison with the interactive segmentation methods in Fig. \ref{fig:inter_pkg}. We can see in the figure that One-Prompt Model outperforms all other interactive competitors by a significant margin. These results demonstrate that One-Prompt Model could perform as well when the query image itself is prompted, despite not being intentionally trained for this specific setting. By training under our more challenging setting, the One-Prompt Model demonstrates superior capability compared to interactive models.

\begin{table*}[h]
\centering
\caption{The zero-shot comparison between One-Prompt Model, full-supervised models and SAM-based models under 'segment everything' setting. Evaluated on 11 unseen tasks by Dice Score (\%). }
\resizebox{0.9\linewidth}{!}{%
\begin{tabular}{c|ccccccccccc|c}
\hline
Methods                                                      & KiTS         & ATLAS            & WBC          & SegRap           &  CrossM          & REFUGE          & Pendal          & LQN            & CAS          & CadVidSet            & ToothFairy                  & Ave            \\ \hline
TransUNet                                                    & 38.2          & 34.5          & 49.1          & 25.5          & 37.7          & 36.3          & 31.2          & 23.3          & 24.5          & 31.6          & 37.9                    & 33.6          \\ 
Swin-UNetr                                                  & 37.2          & 26.5          & 32.1          & 25.6          & 29.7          & 28.9          & 31.4          & 17.2          & 20.5          & 22.6          & 32.1                    & 28.5          \\ 
nnUNet                                                       & 39.8        & 30.3          & 40.4          & 26.8          & 35.0          & 34.9          & 42.9           & 18.9          & 37.4          & 41.8          & 35.3                  & 34.9          \\ 
MedSegDiff                                                   & 40.1          & 30.5          & 42.9          & 34.7          & 37.7          & 31.9          & 42.6          & 21.1          & 38.3          & 34.7          & 33.5                    & 35.3          \\  \hline
\begin{tabular}[c]{@{}c@{}} MSA\end{tabular}  & 54.6          & 48.9          & 55.9          & 47.3          & 51.7          & 49.2          & 54.2          & 41.0          & 48.9          & 53.5          & 47.6                    & 50.3          \\              
MedSAM  & 62.4          & 53.1          & 67.8          & 52.3          & 59.3          & 54.5          & 58.7          & 42.5          & 41.5          & 45.7          & 56.2                    & 53.9          \\ 
SAM-Med2D                                                       & 56.3          & 51.4          & 52.6          & 43.5          & 47.2          & 52.0          & 50.8          & 47.4          & 44.3          & 49.0          & 55.1                    & 50.0                  \\ 
\rowcolor{cyan!40!white!20}
One-Prompt  &  67.3  & 63.8 &  72.5  &  62.2  &  65.8  & 58.4          &  72.6  &  49.5  &  64.5  &  66.3  &  61.4   &  64.0  \\ 
\hline

\end{tabular}%
}\label{tab:amos}
\end{table*}

\subsection{Zero-shot Capability} \label{sec:exp-zeroshot}
\noindent \textbf{Task} Our model can automatically segment all salient targets following the similar 'segment everything' setting in SAM. In this setup, we prompt the template image with a regular grid of foreground points, generating an average of approximately 50 masks per image. We compare our model under this setting with conventional fully-supervised models that are not promptable \cite{chen2021transunet, hatamizadeh2022swin, isensee2021nnu, wu2022medsegdiff, hatamizadeh2022unetr, chen2023transattunet}, and also SAM-based methods \cite{MedSAM, wu2023medical, cheng2023sam} under 'segment everything' setting.

\noindent \textbf{Data}
We use 11 unseen datasets in the held-out test set to verify the zero-shot transfer ability of the models. Detailed datasets are shown in Table. \ref{tab:amos}

\noindent \textbf{Results}
We present a quantitative comparison of zero-shot segmentation results in Table \ref{tab:amos}. It shows the challenge faced by fully-supervised segmentation methods in generalizing to unseen tasks, as they may struggle to understand the task, such as 'what to segment,' without the human interaction. When compared to SAM-based models under the 'segment everything' setting, the One-Prompt Model consistently outperforms them across all tasks, achieving the highest average performance of 64.0\%, which surpasses the second-highest by a substantial 10.7\%. It again highlights the value of setting a challenging learning task with a comparable model for enhancing the generalization.

\subsection{Ablation Study and Analysis} \label{sec:exp-ablation}
\begin{figure}
    \centering
    \includegraphics[width=0.95\linewidth]{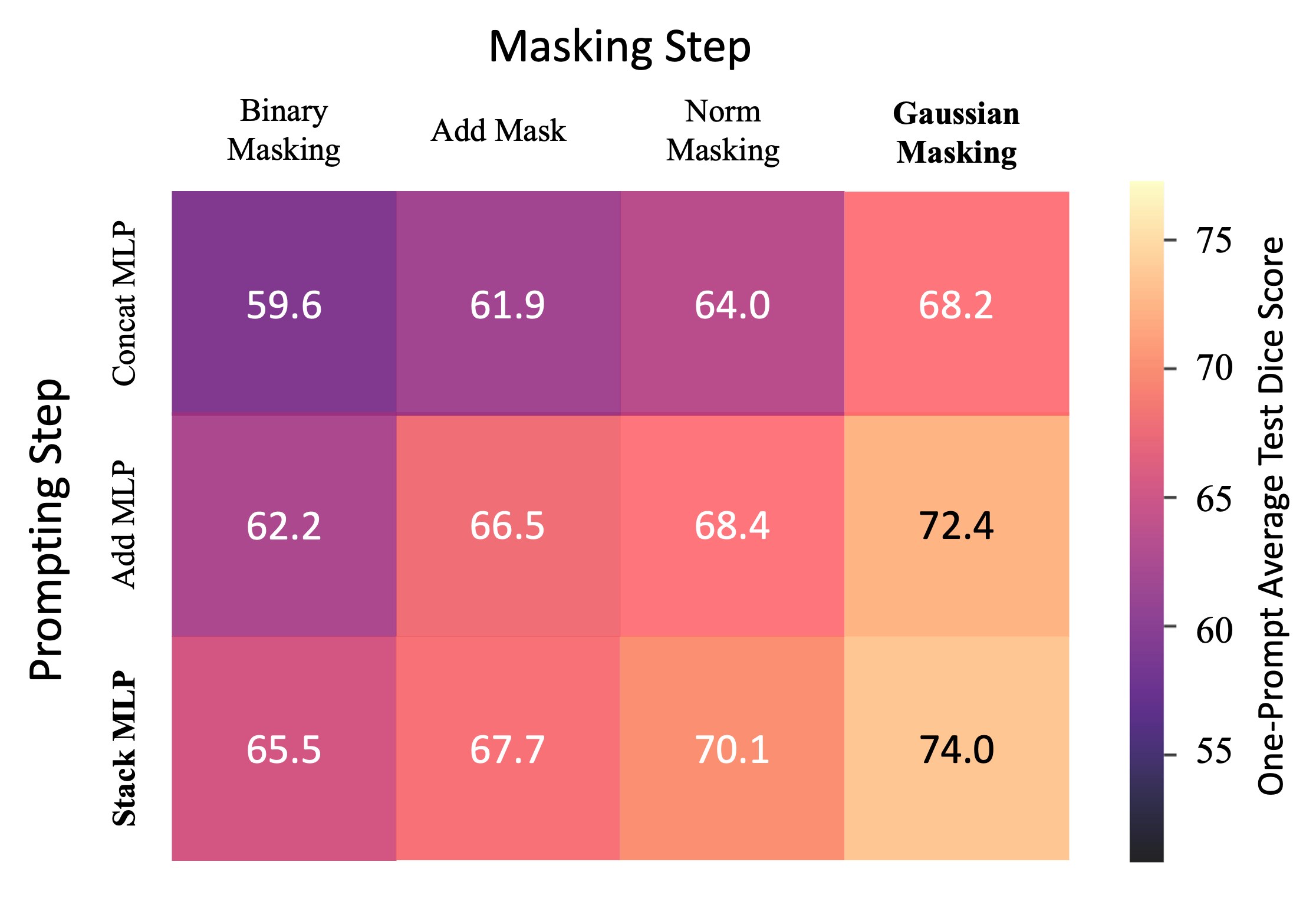}
    \caption{Ablation study on Prompt-Parser. We cross-validate the combination of different methods in Masking and Prompting steps, and show the average dice score under one-prompt segmentation setting on the held-out test set.}
    \label{fig:ablation}
\end{figure}

\noindent \textbf{Ablation on Prompt-Parser}
In the design of Prompt-Parser, we experimented with various combinations of strategies in both the Masking and Prompting steps. The comparative results are shown in Fig. \ref{fig:ablation}. For the Prompting step, we explored simply adding or concatenating three embeddings and then projecting them to the desired length using a MLP. In the Masking step, we tested different approaches, such as directly adding the mask to the feature, binary thresholding the mask (setting negatives to zero and positives to one) then do element-wise multiplication with the feature (\textit{Binary Masking}), or normalizing the mask and then element-wise multiplying it with the feature (\textit{Norm Masking}). We can see the combination of the proposed Stack MLP + Gaussian Masking achieves the highest score on the held-out test dataset.

\begin{figure}
    \centering
    \includegraphics[width=0.95 \linewidth]{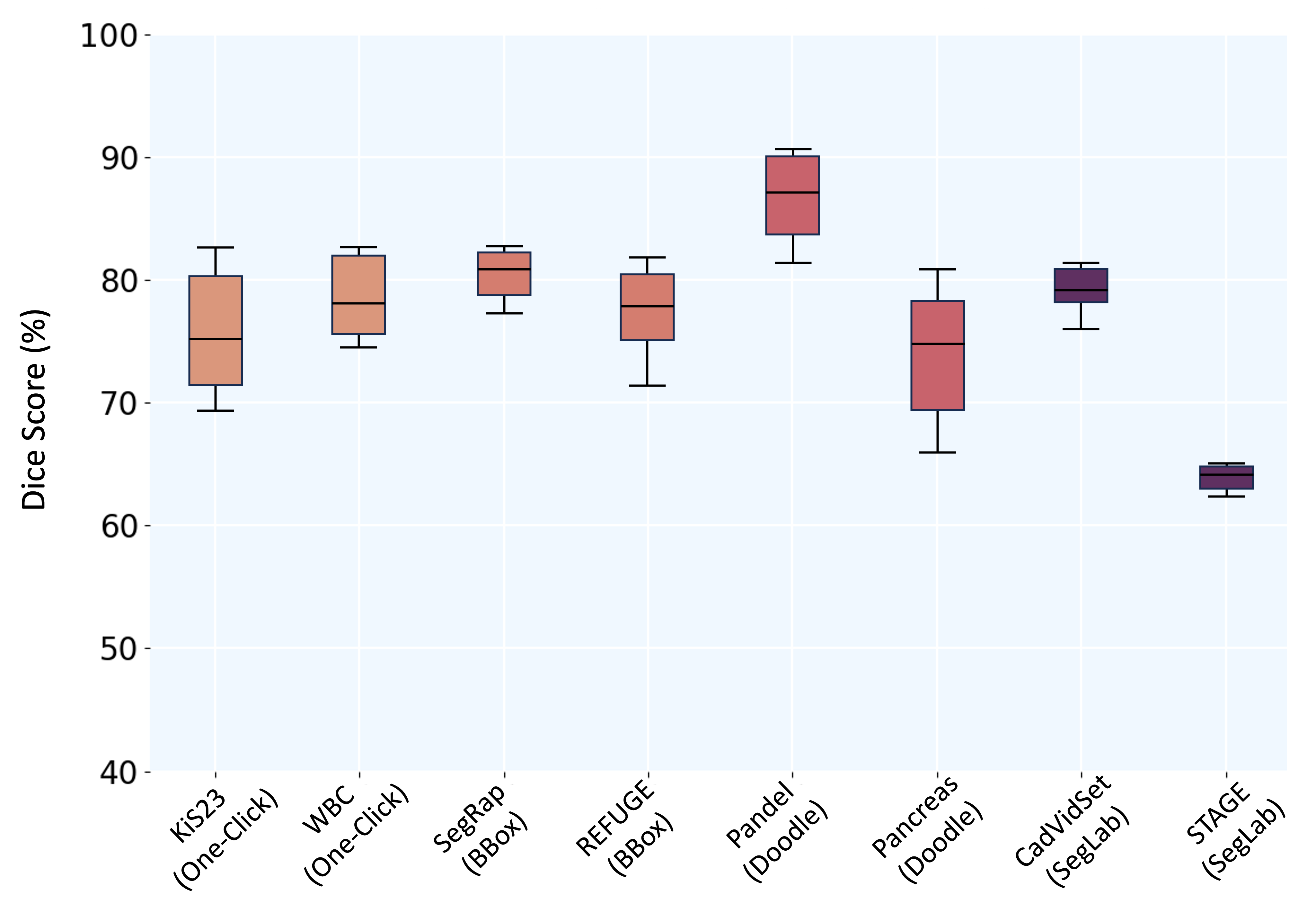}
    \caption{Variance of offering different templates to One-Prompt Model in the inference. Validated on 8 held-out test task giving different prompts.}
    \label{fig:variance}
\end{figure}

\noindent \textbf{Variance of offering different Templates in the inference}
To assess the variance of using different templates during inference, we conducted 100 repetitions using different random templates across 8 test tasks. The results are shown in Fig. \ref{fig:variance}. We can see given the same type of prompts for different tasks, the larger variances are shown in tasks with more diverse and uncertain target structures, such as \textit{Optic-Cup} (REFUGE) and \textit{Pancreas} segmentation. This is because the template may significantly differ from the query samples in these tasks. The variance also varies depending on the type of prompts, such as the notably smaller variance is observed when using fine-grained \textit{SegLab} prompts. Overall, we observed variance consistently staying below 13\%. This inherent stability of the model suggests a robust zero-shot generalization capability.

\noindent \textbf{Effect of prompt quality \& types in the inference}
To verify the effect of prompt quality and types in the inference, we categorize five different levels of the prompt quality, from the lowest to the highest quality, respectively denoted as: \textit{Low}, \textit{Medium}, \textit{High}, \textit{Oracle}, and \textit{Human}, on each of the prompt type. The detailed prompt simulation process is provided in the supplementary. We provide the model the same template with different prompt qualities each time in the comparison. The model is tested under One-Prompt setting on REFUGE and WBC datasets.

\begin{figure}[h]
    \centering
    \includegraphics[width= \linewidth]{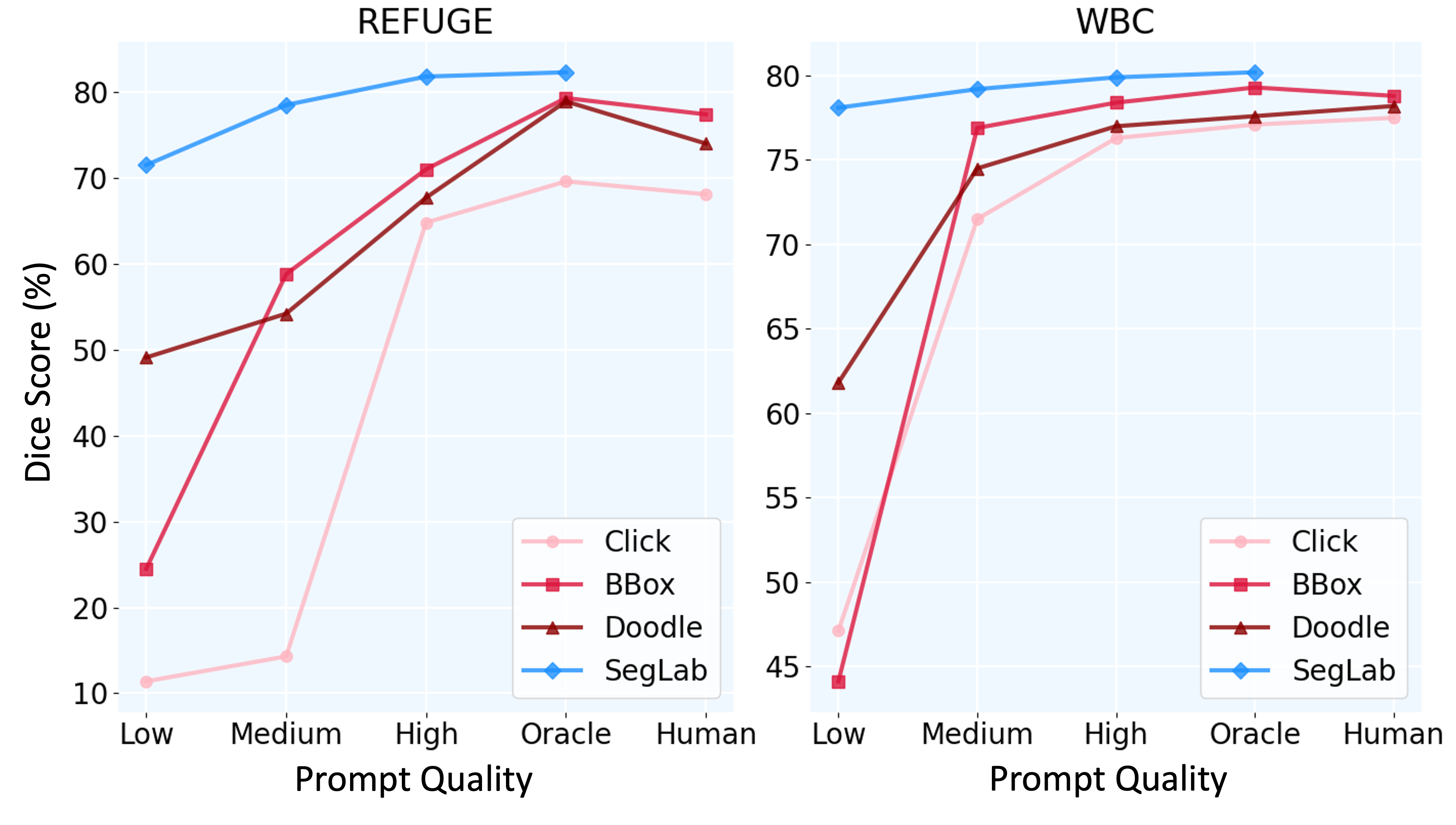}
    \caption{The variance of model performance given different prompts with different qualities on REFUGE and WBC test datasets.}
    \label{fig:prompt-quality}
\end{figure}

In Fig. \ref{fig:prompt-quality}, we observe a gradual improvement in model performance as prompt quality increases, highlighting the significant impact of prompt quality on the final model performance. The choice of prompt types also has a notable effect on the results. For both tasks, fine-grained \textit{SegLab} exhibit the highest performance. \textit{BBox} and \textit{Doodle} demonstrate comparable performance after convergence and generally outperform the quicker and simpler \textit{Click} prompt. This underscores a trade-off between user prompting and model performance: achieving better performance typically requires more detailed and high-quality prompts.

\noindent \textbf{Model Efficiency}
In Table \ref{tab:model_eff}, we compare the efficiency of our model with one/few-shot learning models across 14 test tasks. Additionally, we train 14 TransUNet on 14 test datasets to establish an upper-bound for the performance. Unlike current one/few-shot models that require the fully-labeled images, One-Prompt only needs the users to simply prompt the image, significantly reducing the user-cost time. On average, it takes a user 27.47 seconds to annotate one image across 14 datasets, while prompting for one image only takes an average of 2.28 seconds. Moreover, the prompting process requires the users much less clinical background and domain knowledge, making it more practical. The One-Prompt Model also exhibits superior scale-up capability, showing a significant improvement of about 10\% compared to smaller models and only a 3.23\% decrease compared to the TransUNet upper-bound. Comparing with the fully-supervised upper-bound, One-Prompt only needs to train one time for all downstream tasks, which saves significant parameters, training run time, and user-cost time for the annotation.
 
\begin{table}[h]
\centering
\caption{ \textbf{Model efficiency comparison with few/one-shot transfer learning models.} To establish an upper bound for performance, we individually train 14 task-specific TransUNet models for 14 held-out datasets. The run-time is its cumulative training time. The user-cost time is denoted as $\infty$ since the user must annotate all training samples in using.
}
\resizebox{\linewidth}{!}{%
\begin{tabular}{c|cccc}
\hline
Models                    & params (M)  & Run Time (ms)                    & User-Cost Time (s) & Dice \\ \hline
ALPNet     & 14.7     & 240                        & 27.47                  &  52.96        \\
PANet      & 43.0     & 528                        & 27.47                  &   50.11       \\
HyperSegNas     & 1321.0      & 2154                          & 27.47                  &   63.86       \\
UniverSeg   & 1.2      & 142                        & 27.47                  &  64.66        \\ 
\rowcolor{cyan!40!white!20}
OnePrompt & 192.0     & 741                              & 2.28                   & 73.98         \\ \hline
TransUNet (sup.)              & 14 $\times$ $10^{3}$  & 14 $\times$ 5.7 $\cdot 10^{7}$                        & $\infty$                      &   77.21       
\end{tabular}} \label{tab:model_eff}
\end{table}

\section{Conclusion}
In this paper, we introduce "One-Prompt Medical Image Segmentation", a new paradigm for building foundation model to handle diverse medical segmentation tasks. The model competed many related methods with just one prompted sample. With user-friendly prompt options for clinicians and remarkable results, the model holds significant promise for practical applications in clinical settings.

\clearpage
\clearpage
{
    \small
    \bibliographystyle{ieeenat_fullname}
    \bibliography{main}
}


\end{document}